\documentclass[aps,prb,msmath,amssymb,amsfonts,superscriptaddress,floatfix]{revtex4} 
\pdfoutput=1
\usepackage{amsmath,amsthm,amsfonts,amssymb}
\usepackage{graphicx}

\usepackage{color}

\newtheorem{theorem}{Theorem}

\newtheorem{lemma}{Lemma}

\usepackage{algorithm}

\DeclareMathAlphabet{\mathpzc}{OT1}{pzc}{m}{it}

\newcommand{\en}{S}

\newcommand{\be}{\begin{equation}}
\newcommand{\ee}{\end{equation}}

\newcommand{\pplus}{\psi_{+}}
\newcommand{\expec}{\mathbb{E}}
\newcommand{\sO}{{\mathcal O}^*}
\newcommand{\mO}{{\mathcal O}}
\newcommand{\poly}{{\rm poly}}
\newcommand{\sz}{S^{(comp)}}

\newcommand{\Po}{P_{ov}}

\newcommand{\cfn}{\tau}

\newcommand{\bex}{\mu}
\newcommand{\mA}{\hat{A}}
\newcommand{\mB}{\hat{B}}
\newcommand{\mC}{\hat{C}}
\newcommand{\mAp}{\hat{A}'}

\begin{document}

\title{Weaker Assumptions for the Short Path Optimization Algorithm}

\author{Matthew B.~Hastings}

\affiliation{Station Q, Microsoft Research, Santa Barbara, CA 93106-6105, USA}
\affiliation{Quantum Architectures and Computation Group, Microsoft Research, Redmond, WA 98052, USA}
\begin{abstract}
The short path algorithm\cite{spa} gives a super-Grover speedup for various optimization problems under the assumption of a unique ground state and under an assumption on the density of low-energy states.  Here, we remove the assumption of a unique ground state; this uses the same algorithm but a slightly different analysis and holds for arbitrary MAX-$D$-LIN-$2$ problems.  Then, specializing to the case $D=2$, we show that for certain values of the objective function we can always achieve a super-Grover speedup (albeit a very slight one) without any assumptions on the density of states.  Finally, for random instances, we give a heuristic treatment suggesting a more significant improvement.
\end{abstract}
\maketitle

\section{Introduction}
The short path algorithm\cite{spa} is a quantum algorithm for exact combinatorial optimization.  One is given some Hamiltonian
$H_Z$ that is diagonal in the computational basis in a system on $N$ qubits.  Then, one prepares an initial state $\pplus$
which is maximally polarized in the $X$ basis, and one phase estimates this state using a Hamiltonian $H_1$ which is equal to $H_Z$ plus a certain off-diagonal perturbation.  If the phase estimation succeeds in giving the ground of $H_1$, then the resulting state has a large squared overlap with the ground state of $H_Z$.  Here, ``large" means that the squared overlap is significantly larger than $2^{-N}$, even if it is possibly still exponentially small.
In this sense, the algorithm uses a ``short path": the ground state of $H_1$ is close to that of $H_Z$, so that one only slightly changes the Hamiltonian rather than trying to follow an adiabatic path\cite{farhi2001quantum} from a Hamiltonian whose ground state is $\pplus$ to one whose ground state is a ground state of $H_Z$.

The idea of the short path algorithm is that it allows us to sidestep one of the fundamental problems with the adiabatic algorithm, that gaps may become superexponentially small along the adiabatic path\cite{altshuler2010anderson}.  Indeed, while specific examples can show very pathological behavior of the gaps\cite{wecker2016training}, the argument of
Ref.~\onlinecite{altshuler2010anderson} is that this problems of small gaps is inevitable due to many-body localization: those authors claim that for many choices of $H_Z$, the ground state wavefunction for any small perturbation will be localized to a small region of the Boolean hypercube and this will lead to (avoided) level crossings if some wavefunction supported in some other region of the hypercube has an energy that becomes close to the ground state energy. For many choices of perturbation, these crossings will be avoided but will still give very small gaps.
The idea of the short path algorithm then is accept that the ground state wave function of $H_1$ may be localized, but to take advantage of tails of the wavefunction that spread across the hypercube; these tails can be exponentially small and still lead to a nontrivial speedup.  One may say that the idea is that the ground state wavefunction may be localized using an $\ell_2$ norm but might be delocalized using an $\ell_1$ norm; that is, in the computational basis, the squared amplitudes may be localized near some corner of the Boolean hypercube, but the absolute value of the amplitude may be delocalized across the hypercube.

This tail is how the algorithm achieves a speedup compared to Grover search\cite{grover1996fast}.  The delocalized wavefunction (using the $\ell_1$ norm) implies that the $\ell_1$ norm of the wavefunction may be much larger than $1$ even if the $\ell_2$ norm is normalized to $1$.  The Hamiltonian is constructed so that all coefficients of the ground state wavefunction have the same sign, so that the $\ell_1$ norm of the ground
state wavefunction is equal to $2^{N/2}$ times its inner product with the state $\pplus$; hence, 
this large $\ell_1$ norm implies a large inner product with the state $\pplus$.

The amount of speedup depends upon some parameters in the algorithm, and in turn the success of the algorithm requires some relationship between these parameters and the number of computational basis states with low
energy for $H_Z$.  Thus, the speedup is not unconditional, but rather requires some promises.
One promise that we remove for all instances is the requirement of a unique ground state of $H_Z$; this is done using a more careful analysis of the algorithm that also holds for $H_Z$ with degenerate ground states.
We also remove the remaining promises, which involved the density of states, for certain values of the objective function to give a very slight unconditional speedup that we term a ``mini-super-Grover" speedup (however, giving additional promises on the density of states gives a more significant speedup as before).
We emphasize that no classical algorithm is known for MAX-$D$-LIN-$2$ that is faster than $\sO(2^N)$,
where
$\sO(\ldots)$ is a soft-O notation, indicating a bound up to polylogarithms, which in this case are polynomials in $N$, other than the case $D=2$ where such an algorithm is known but
requires exponential space\cite{williams2005new}.

\subsection{Problem Statement and Summary of Results}
As in Ref.~\onlinecite{spa}, in this paper we let $H_Z$ be any Hamiltonian that is a weighted sum of products of Pauli $Z$ operators, 
with each product containing exactly $D$ such operators on distinct qubits for some given $D$.  

One minor generalization compared to Ref.~\onlinecite{spa} is
that each product has a weight that may be an arbitrary real number, while in Ref.~\onlinecite{spa} we required all weights to be integer.  We define $J_{tot}$ to be the sum of the absolute values of the weights.  We require that $J_{tot}=\mO({\rm poly}(N))$.  Indeed, we fix any $\beta>0$ and require $J_{tot}=\mO(N^\beta)$.  (If all weights are chosen from $\{-1,+1\}$ we have $J_{tot}=\mO(N^2)$.)

Let $E_0$ denote the ground state energy of $H_Z$.  We consider the problem of finding a computational basis state which is a ground state of $H_Z$, assuming that $E_0$ is known,
and assuming that all there is a gap of at least $1$ between the ground state and first excited state of $H_Z$, i.e., that all eigenvalues of $H_Z$ are either equal to $E_0$ or greater than or equal to $E_0+1$.
This gap assumption holds automatically if all weights are integers (as in Ref.~\onlinecite{spa}) since then all eigenvalues are integers.
In fact, the only place that the integer weights are used in Ref.~\onlinecite{spa} is to give the gap assumption and in a certain use of ``binning" in the entropy calculation described later.

There are three main results in this paper.
The first result is that we remove the requirement that the ground state of $H_Z$ be non-degenerate, as well as making some other, less important, generalizations to the result. This requirement of a non-degenerate ground state in Ref.~\onlinecite{spa} was due to how the algorithm was analyzed.  Here, we do not change the algorithm, but we give a more careful analysis that shows that this requirement is not necessary; some slight changes to subleading terms in the speedup of the algorithm occur as a result.  The different analysis here uses a modification of the Brillouin-Wigner perturbation theory that we describe later.  The results here are in theorem \ref{Qgood} which generalizes theorem 3 of Ref.~\onlinecite{spa} and theorem \ref{mainconst} which generalizes theorem 2 of Ref.~\onlinecite{spa}.

The second result is for the case $D=2$.  In this case we show a small but unconditional super-Grover speedup for certain values of the objective function.
This is in section \ref{arbitrary}.

Finally, in section \ref{random} we give physics arguments for why an even larger speedup is expected for random models; in contrast to the rest of the paper, the results in section \ref{random} are just heuristic arguments.

\section{Degenerate Ground States}
\label{dgs}
Here we present the analysis in the case of degenerate ground states.
In subsection \ref{ar}, we review the algorithm.  
In subsection \ref{res} we give the main results of this section, showing a speedup for the algorithm.
In subsection \ref{mbwpt} we present a modification of the Brillouin-Wigner perturbation theory to analyze the algorithm.  In subsection \ref{co} we address convergence of this perturbation theory and compute an overlap needed to estimate the speedup.  In subsection \ref{pt} we prove theorem \ref{Qgood} below.
In subsection \ref{mainproof} we  prove theorem \ref{mainconst} below.

First we need some definitions. 

Assume that $H_Z$ has some number, $n_0$, of degenerate ground states with eigenvalue $E_0$.
All other eigenvalues are at least $E_0+1$.
Let $P$ project onto the ground state subspace of $H_0$ and let
$Q=1-P$.

Let $H_s=H_Z-sB(X/N)^K$.
Unless otherwise specified, whenever we mention $H_s$ we assume that $s\in [0,1]$.
We always assume that $B>0, B=\mO({\rm poly}(N))$.

Consider the Hamiltonian $Q H_s Q$.  Let $E^{Q}_{0,s}$ be the smallest eigenvalue of this Hamiltonian in the subspace spanned by the range of $Q$. 

All logarithms in this paper are to base $2$ unless otherwise stated.

\subsection{Algorithm Review}
\label{ar}
Here we review the algorithm.  We will give several different assumptions in this subsection; the later subsections will then show conditions under which these assumptions hold.

Let $P_s$ project onto the $n_0$ lowest energy states of $H_s$, so that $P_0=P$.
For $s>0$ and $K$ odd the ground state is unique by Perron-Frobenius (as explained in Ref.~\onlinecite{spa}, the Hamiltonian is irreducible on the computational basis for $K$ odd).
For $K$ odd, 
let $\psi_{0,s}$ be the ground state wavefunction of $H_s$.

In this paper, we will allow $K$ to be even or odd, generalizing Ref.~\onlinecite{spa} where $K$ needed to be odd. 
For $s>0$ and $K$ even, the ground state may be non-unique.  For $K$ even, the Hamiltonian is reducible in the computational basis; it is block diagonal consisting of two blocks, one containing the even
computational basis states and one containing the odd computational basis states, where the ``even" computational basis states are those with even Hamming weight, i.e., they are $+1$ eigenstates of $Z_1 Z_2 \ldots Z_N$, and the ``odd" computational basis states are those with odd Hamming weight.
However, within each of those two blocks, the Hamiltonian is irreducible for $K$ even.
For $K$ even, choose one of these blocks (i.e., choose either the even or odd computational basis states) such that a ground state of
$H_Z$ is in that block (if both blocks contain a ground state of $H_Z$, the choice may be done arbitrarily), and let $\psi_{0,s}$
be the ground state wave function of $H_s$ in that block; by Perron-Frobenius, this $\psi_{0,s}$ is unique.
Whenever we consider even $K$ later, we will restrict to that block.

We normalize $|\psi_{0,s}|=1$.
Let $E_{0,s}$ be the ground state energy of $H_s$.

Let us assume that the largest eigenvalue of $H_1$ in the range of $P_1$ is at most $E_0+1/4$ and
let us assume that the smallest eigenvalue of $H_1$ in the range of $1-P_1$ is at least $E_0+1/2$.
Later, we will give conditions under which this holds.

We use the same algorithm as in Ref.~\onlinecite{spa}.
That is, we apply amplitude amplification\cite{brassard2002quantum} to the algorithm \ref{spa2} below.
We perform the phase estimation in step {\bf 2} to precision smaller than $1/8$ so that we can distinguish the case that the energy is at most $E_0+1/4$ from the case that the energy is at least $E_0+1/2$, i.e., so that we can accurately distinguish the $n_0$ smallest eigenvalues from the rest of the spectrum.
We use the same phase estimation procedure as in Ref.~\onlinecite{spa}.  As in Ref.~\onlinecite{spa}, for $J_{tot}=\mO({\rm poly}(N))$, the phase estimation procedure can be carried out to error $\epsilon$ in time polynomial in $N$ and polylogarithmic in $\epsilon$, so we can take $\epsilon$ exponentially small in $N$ and still perform the algorithm in time polynomial in $N$.

\begin{algorithm}
\caption{Simplified Short-Path (unamplified version)}
\begin{itemize}
\item[{\bf 1.}] Let $\psi=\pplus$ be the input state, where
$\pplus=|+\rangle^{\otimes N}$.
\item[{\bf 2.}] Phase estimate $\psi$ using Hamiltonian $H_{1}$.
If the energy estimate is greater than $E_0+1/2$, then terminate the algorithm and return failure. 
\item[{\bf 3.}] Measure the state in the computational basis and compute the value of $H_Z$ after measuring.
If this value is equal to $E_0$ then declare success and output the computational basis state.
\end{itemize}
\label{spa2} 
\end{algorithm}

Let $\Po=\langle \pplus| P_1 | \pplus\rangle $.
Thus, the probability that the measurement in step {\bf 2} succeeds is
$\Po-\mO(\epsilon)$.
Later, we will give conditions under which for any vector $\psi$ in the range of $P_1$ we have that
\be
\label{1meas}
\langle \psi | P_0 | \psi \rangle = \Omega(1).
\ee
Thus, under this assumption
the quantum algorithm \ref{spa2} succeeds with probability at least 
$\Omega(1) \cdot \Po-\mO(1) \cdot \epsilon$
 in finding the ground state of $H_Z$.
We take $\epsilon$ sufficiently small compare to $2^{-N}$ so that the error $\epsilon$ is negligible compared to $\Po$ computed later.
Hence,
applying the method of amplitude amplification to algorithm \ref{spa2},
one obtains an algorithm which succeeds in producing the ground state of $H_Z$ in expected time 
$\mO(\Po^{-1/2}){\rm poly}(N)$.

Note that
\be
\Po \geq |\langle \psi_{0,1} | \pplus \rangle|^2.
\ee
We will lower bound $|\langle \psi_{0,1} | \pplus \rangle|^2$ later which lower bounds $\langle \pplus| P_1 | \pplus\rangle$.

\subsection{Results}
\label{res}
The main result regarding the algorithm that we prove is the following theorem which
 generalizes theorem 2 of Ref.~\onlinecite{spa}.
 The second possibility, item 2 in the theorem, involves the existence of a large number of low energy states.  Commonly in physics, the number of such states of a given energy is referred to as a density of states, and so we refer to a ``density of states assumption" later when we mean an assumption that item 2 of the theorem cannot hold (and hence that the speedup of item 1 must hold).
 
\begin{theorem}
\label{mainconst}
Let $B=-bE_0$ with $0\leq b <1$.  Let $K$ be a positive integer.
Assume $B=\omega(\log(N))$.
Let $W(E)$ be the number of computational basis states which are eigenstates of $H_Z$ with energy in the interval $[E,E+1)$.

Assume that $H_Z$ has $n_0$ ground states.
Assume that
\be
\label{Kbound}
B \cdot \cfn\Bigl(\frac{\log(n_0)}{N}+\frac{(K+1/2)\log(N)+1}{N}\Bigr)^{K} \leq 1/4.
\ee

Then, at least one of the following holds:
\begin{itemize}
\item[1.] The algorithm finds the ground state in expected time
$$\sO\Bigl(2^{N/2} \exp\Bigl[-\frac{b}{2DK} N    \cdot (1-o(1))    \Bigr]\Bigr).$$

\item[2.] For some $X_0\geq X_{min}=N\cdot (10B)^{-1/K}$,
there is some probability distribution $p(u)$ on computational basis states with entropy in bits at least
$$\sz\geq N \cfn^{-1}\Bigl(\frac{X_0-X_{min}/K}{N}\Bigr)$$ 
and with expected value of $H_Z$
at most
$$E_0+\mO(1) \frac{J_{tot} K^2 D^2}{X_{min}^2}+\frac{5}{2} B\Bigl(\frac{X_0+X_{min}/K}{N}\Bigr)^K\cdot \mO(1).$$
Further, for some function $F(S)$ with
\be
\label{FSdef}
F(S)=
E_0+\mO(1) \frac{J_{tot} K^2 D^2}{X_{min}^2}+\frac{5}{2}B(\cfn(S/N))^K\cdot \mO(1),
\ee
then for some integer $E>E_0$ we have $\log(W(E)) \geq F^{-1}(E)-\mO(\log(N))$.
\end{itemize}
\end{theorem}
The most important change compared to theorem 2 of Ref.~\onlinecite{spa} is that the degeneracy assumption has been
removed.
The function $\cfn(\cdot)$ in the theorem is a continuous increasing function, taking $[0,1]$ to $[0,1]$.
For small $\sigma$,
\be
\cfn(\sigma)=\Theta(\sqrt{\frac{\sigma}{-\log(\sigma)}}).
\ee

The assumption that $B=\omega(\log(N))$ is very mild: this holds for fixed $b$ if $|E_0|=\omega(\log(N))$ which is a very mild assumption; this assumption was not needed in Ref.~\onlinecite{spa} because there the unique ground state assumption meant that $|E_0|=\Omega(N))$.
 Indeed if all qubits participate in at least one term in $H_Z$ with nonzero weight and all weights are integers, then $|E_0|=\Omega(\sqrt{N})$ by results in Ref.~\onlinecite{haastad2004advantage}.

To understand the assumption in Eq.~(\ref{Kbound}), note that $$\frac{(K+1/2)\log(N)+1}{N}=\mO(\log(N)/N)$$ for any fixed $K$.
Hence, if $\log(n_0)=\mO(N^\alpha)$ for any $\alpha<1$, we have
$$\cfn\Bigl(\frac{\log(n_0)}{N}+\frac{(K+1/2)\log(N)+1}{N}\Bigr)=\mO(N^{(\alpha-1)/2}/\log(N))$$ and hence for
$B=o(N^{K(\alpha-1)/2})$, this assumption will be satisfied for sufficiently large $N$.
Even if $\log(n_0)\leq aN$ for any constant $a<1$, the assumption will be satisfied for $K=\Omega(\log(N))$ for $B=\mO(\poly(N))$.

A minor change is that $K$ does not need to be odd.  The requirement of odd $K$ in Ref.~\onlinecite{spa} occurred in three places.  The
first place was to make the ground state of $H_s$ unique for $s>0$ by Perron-Frobenius; we deal with this for even $K$ by choosing a block (either even or odd computational basis states) in which $H_s$ is irreducible and working within that block so that the ground state of $H_s$ is unique in that block as explained above.
The second place was to make $\langle 0 | (X/N)^K | 0\rangle$ vanish for any computational basis state $|0\rangle$.
This requirement is not necessary here due to more careful treatment of the matrix elements of $(X/N)^K$ between ground states and is replaced by assumption \ref{Kbound}.  The third place involved the log-Sobolev inequalities and localization in $X$; we explain how this part of the proof is modified in section \ref{mainproof}.

The previous subsection defined the algorithm, and
gave several conditions to estimate the success probability of the algorithm.
We needed to show an upper bound $E_0+1/4$ on the largest eigenvalue of $H_1$ in the range of $P_1$ and a lower bound $E_0+1/2$ on the smallest eigenvalue outside this range; we needed to
lower bound 
$|\langle \psi_{0,1} | \pplus \rangle|^2$;
and we needed to show Eq.~(\ref{1meas}).

We now summarize our results for these quantities in the form of the following theorem which generalizes theorem 3 of Ref.~\onlinecite{spa}. 
 These results are proven in later subsections.
 Theorem \ref{Qgood} will be used to show theorem \ref{mainconst}; we will show later that the assumptions of theorem \ref{Qgood} hold given the assumptions of theorem \ref{mainconst} and given an assumption that item 2. of theorem \ref{mainconst} does not hold.
\begin{theorem}
\label{Qgood}
Consider the Hamiltonian $Q H_s Q$.  Let $E^{Q}_{0,s}$ be the smallest eigenvalue of this Hamiltonian in the subspace spanned by the range of $Q$. Assume that $E^{Q}_{0,1}\geq E_0+1/2$ and assume that
$\Vert P B(X/N)^K \Vert \leq 1/4$.
Assume $B=\omega(\log(N))$.

Then
\begin{itemize}
\item[1.] There is an upper bound $E_0+1/4$ on the largest eigenvalue of $H_1$ in the range of $P_1$ and a lower bound $E_0+1/2$ on the smallest eigenvalue of $H_1$ outside this range.

\item[2.] 
We have
\begin{eqnarray}
\langle \pplus | \psi_{0,1} \rangle & \geq &
\Omega(1) \cdot 2^{-N/2} \exp\Bigl(\frac{BN}{2DK|E_0|} \cdot (1-o(1) \Bigr).
\end{eqnarray}

\item[3.] Eq.~(\ref{1meas}) holds.
\end{itemize}

Hence, from items 1-3 above, it follows that: given the assumptions of the theorem, the algorithm of section \ref{ar} succeeds in finding a ground state in
expected time
$\sO(2^{-N/2} \exp\Bigl(\frac{BN}{2DK|E_0|} \cdot (1-o(1) \Bigr))$.
\end{theorem}

The assumptions of theorem \ref{Qgood} include a bound on $\Vert P B(X/N)^K \Vert$.  
Note that $\Vert P(X/N)^K \Vert$ is equal to the maximum of
$\Bigl| (X/N)^{K} | \psi \rangle \Bigr|$ over states $\psi$ with $|\psi|=1$ and $\psi$ supported on the ground states of $H_0$.
In appendix \ref{glsi}, we bound this quantity in terms of $n_0$ by
\be
\label{XNKbnd}
\Bigl| (X/N)^{K} | \psi \rangle \Bigr| \leq \cfn\Bigl(\frac{\log(n_0)}{N}+\frac{(K+1/2)\log(N)+1}{N}\Bigr)^{K},
\ee
so that if Eq.~(\ref{Kbound}) holds then $\Vert P B(X/N)^K \Vert \leq 1/4$.

\subsection{Modified Brillouin-Wigner Perturbation Theory}
\label{mbwpt}
As in Ref.~\onlinecite{spa}, we use a Brillouin-Wigner perturbation theory.  We make two modifications to the perturbation theory.
First, we consider the case that $H_0$ has a degenerate ground state, rather than a unique ground state.
This case is well studied in the literature.  See for example Ref.~\onlinecite{leinaas1978convergence}.
Second, we will modify this perturbation theory in a way explained below to help simplify the treatment of certain ``returning paths" as explained later.

The results in this subsection do not use any properties of the specific choice of $H_s$ above;
we consider a Hamiltonian $H_s=H_0+sV$ in this subsection with $H_0,V$ arbitrary.
Let the ground state subspace of $H_0$ have eigenvalue $E_0$.

Let
\be
G_0(\omega)=(Q (\omega-H_0) Q)^{-1},
\ee
where $\omega$ is a scalar and
where the inverse is computed in the subspace which is
the range of $Q$ and let $(1-Q)G_0=G_0(1-Q)=0$.
That is, $G_0(\omega)$ is a Moore-Penrose pseudo-inverse of $Q (\omega-H_0)) Q$, so
that $G_0(\omega) (\omega-H_0)=(\omega-H_0)G_0(\omega)=Q$.

Define an ``effective Hamiltonian"
\be
h(\omega,s)=PH_0P + PsV\sum_{k\geq 0} \Bigl(sG_0(\omega) V\Bigr)^kP.
\ee
Let $\xi$ be an eigenfunction of this Hamiltonian with $\xi$ in the range of $P$ and with $$h(\omega,s) \xi=\omega \xi.$$
Note that this means that $\omega$ must be computed self-consistently: the effective Hamiltonian $h(\omega,s)$ depends upon $\omega$.

Then,
let
\be
\label{phidef}
\Phi_s=\xi+
\sum_{k\geq 1} \Bigl(sG_0(\omega) V\Bigr)^k \xi.
\ee

One may then verify as a formal power series that
\be
H_s \Phi_s = \omega \Phi_s,
\ee
so that $\Phi_s$ is an eigenvector of $H_s$ with eigenvalue $\omega$.
To see this,
note that
\begin{eqnarray}
&&\Bigl((H_0-\omega)+sV\Bigr) \Phi_s \\ \nonumber
&=&
(H_0-\omega)\xi + sV\xi
+\Bigl((H_0-\omega)+sV\Bigr) \sum_{k\geq 1} \Bigl(sG_0(\omega) V\Bigr)^k \xi.
\\ \nonumber
&=&
(H_0-\omega)\xi + sV\xi
-\sum_{k\geq 0} sQV \Bigl(sG_0(\omega)V\Bigr)^k\xi+\sum_{k\geq 1} sV \Bigl(sG_0(\omega)) V\Bigr)^k \xi
\\ \nonumber
&=&
(H_0-\omega)\xi
-\sum_{k\geq 0} sQV \Bigl(sG_0(\omega)V\Bigr)^k\xi+\sum_{k\geq 0} sV \Bigl(sG_0(\omega)) V\Bigr)^k \xi
\\ \nonumber
&=&
(H_0-\omega)\xi
+P\sum_{k\geq 0} sV \Bigl(sG_0(\omega)) V\Bigr)^k \xi
\\ \nonumber
&=& (h(\omega,s)-\omega) \xi.
\end{eqnarray}

The above equation holds as a formal power series.
We address convergence later using specific properties of $H_0,V$.

We now describe a simple modification of this perturbation theory.  The reason for this modification is that it will simplify our computation later of overlaps using this perturbation theory; this modification will simplify the treatment because the operator $Q$ will not be in the series.

To motivate this modification, suppose first we considered the series
\be
\label{trial}
``\Phi_s"=\xi+
\sum_{k\geq 1} \Bigl(s(\omega-H_0)^{-1} V\Bigr)^k \xi.
\ee
All manipulations of this series for $``\Phi_s"$ will be formal manipulations of power-series and are intended only as motivation.
Then, we have
$$``\Phi_s"=\sum_{k\geq 0} \Bigl(s(\omega-H_0)^{-1} V\Bigr)^k \xi=(\omega-H_s)^{-1} (\omega-H_0) \xi.$$
Hence, if $\omega$ is an eigenvalue of $H_s$, the inverse $(\omega-H_s)^{-1}$ is not well-defined and the power series (\ref{trial}) is not convergent in general.

However, let us instead define
\be
J_0=H_0+\zeta P,
\ee
where we will take $\zeta>0$ later.
Now define
\be
\label{phiseries}
\phi_s=\sum_{k\geq 0} \Bigl(s(\omega-J_0)^{-1} V\Bigr)^k \xi.
\ee
We now verify as a formal power series that
\be
H_s \phi_s = \omega \phi_s,
\ee
so that $\phi_s$ is an eigenvector of $H_s$ with eigenvalue $\omega$.
We have
\begin{eqnarray}
\phi_s &=&\sum_{k\geq 0} \Bigl(s(\omega-J_0)^{-1} V\Bigr)^k \xi \\ \nonumber
&=&
\xi+\sum_{k\geq 1} \Bigl(sG_0(\omega) V\Bigr)^k \xi 
\\ \nonumber
&&+ \sum_{l\geq 0}  \Bigl(s(\omega-J_0)^{-1} V\Bigr)^l (\omega-J_0)^{-1}
P sV  \sum_{k\geq 0}  \Bigl(s G_0(\omega) V\Bigr)^k  \xi.
\\ \nonumber
&=& \Phi_s+\sum_{l\geq 0}  \Bigl(s(\omega-J_0)^{-1} V\Bigr)^l (\omega-J_0)^{-1}
P sV 
\sum_{k\geq 0} \Bigl(s G_0(\omega) V\Bigr)^k \xi.
\end{eqnarray}
However, by assumption that $\xi$ is an eigenvector of $h(\omega,s)$ with eigenvalue $\omega$, we have that
$P sV  \sum_{k\geq 0} \Bigl(s G_0(\omega) V\Bigr)^k \xi=(\omega-H_0)\xi=(\omega-E_0)\xi$ so
\begin{eqnarray}
\phi_s &=& \Phi_s+\sum_{l\geq 0}  \Bigl(s(\omega-J_0)^{-1} V\Bigr)^l
\frac{\omega-E_0}{\omega-E_0-\zeta} \xi \\ \nonumber
&=& \Phi_s+\frac{\omega-E_0}{\omega-E_0-\zeta} 
\sum_{l\geq 0}  \Bigl(s(\omega-J_0)^{-1} V\Bigr)^l
\xi 
\\ \nonumber
&=& \Phi_s+\frac{\omega-E_0}{\omega-E_0-\zeta} \phi_s,
\end{eqnarray}
so
\be
\phi_s=\frac{\xi+E_0-\omega}{\xi}\Phi_s
\ee
so that $\phi_s$ is equal to $\Phi_s$ times a scalar and hence if $\phi_s$ is nonzero then $\phi_s$ is an eigenvector of $H_s$ with eigenvalue $\omega$.
For the specific choices of $H_0,V,\xi,\zeta$ we choose later, it will be obvious that $\phi_s$ is nonzero since it will be a sum of positive terms in the computational basis.

\subsection{Convergence and Overlap}
\label{co}
In this subsection, we prove the following:
\begin{lemma}
\label{ovlemma}
Suppose that $E_{0,1}\geq E_0-1/2$.
Then,
\begin{eqnarray}
\label{oveq}
\langle \pplus | \phi_{0,1} \rangle 
&\geq &
2^{-N/2} \exp\Bigl[\frac{BN}{2DK|E_0|} \cdot (1-o(1)) \Bigr].
\end{eqnarray}
\end{lemma}

From here on, we set $\omega=E_{0,s}$.
Let $\xi_0$ be the ground state of $h(\omega,s)$, with $|\xi_0|=1$.
For $s>0$, we can choose $\xi_0$ to have all coefficients positive in the computational basis by Perron-Frobenius (recall that for even $K$ we work in a block of either even or odd computational basis states so that $H_s$ is irreducible).

Let $\phi_{0,s}$ be given by
\be
\label{phigs}
\phi_{0,s}=\sum_{k\geq 0} \Bigl(s(\omega-J_0)^{-1} V\Bigr)^k \xi_0.
\ee
That is, $\phi_{0,s}$ is equal to $\phi_s$ if $\xi=\xi_0$ in Eq.~(\ref{phiseries}).

The series (\ref{phiseries}) converges for $|s|\leq 1$, $\omega=E_{0,s}$, $H_0=H_Z$, $V=-B(X/N)^K$ and $\zeta>0$.
To see this, note that it is given by the series expansion of $(\omega-J_0-sV)^{-1} \xi_0$.  Following Ref.~\onlinecite{spa}, singularities of $(\omega-J_0-sV)^{-1} \xi_0$ are simple poles
at values of $s$ such that $J_0+sV$ has an eigenvalue $\omega$, and by Pringsheim's theorem, the closest such singularity since all coefficients of the series have the same sign occurs for $s$ on the positive real axis, i.e., the radius of convergence of the series
is given by the smallest positive real $s$ such that $J_0+sV$ has an eigenvalue equal to $\omega$.
Since the smallest eigenvalue of $J_0+sV$ is larger than $E_{0,s}$ for all $s\in[0,1]$ (this follows by Perron-Frobenius), the claimed convergence holds.

Let us pick $$\zeta=1/2.$$

From Eq.~(\ref{phigs}),
\begin{eqnarray}
\label{rwseries}
\langle \pplus | \phi_{0,1} \rangle &=& \sum_u \langle \pplus | u \rangle \langle u | \xi_0\rangle \\ \nonumber
&&+B\sum_{u,v} \langle \pplus | v \rangle \frac{\langle v | (X/N)^K | u \rangle }{E'_v-E_{0,1}}  \langle u | \xi_0\rangle \\ \nonumber
&&+B^2 \sum_{u,v,w} \langle \pplus | w \rangle \frac{\langle w | (X/N)^K | v \rangle }{E'_w-E_{0,1}} \frac{\langle v | (X/N)^K | u \rangle}{E'_v-E_{0,1}}\langle u | \xi_0\rangle \\ \nonumber
&&+\ldots
\end{eqnarray}
Here, $u,v,\ldots$ label basis states in the computational basis
and $E'_u$ is equal to $\langle u | J_0 | u \rangle$.
For any basis state $u$, we have $\langle \pplus| u \rangle=2^{-N/2}$.

As in Ref.~\onlinecite{spa}, we re-express 
 the series in terms of a random walk on the basis states $|u\rangle$ as follows.  
Let
\be
|\xi_0|_1=\sum_u\langle u | \xi_0\rangle.
\ee
That is, $|\xi_0|_1$ is the $\ell_1$ norm of $\xi_0$; we have that $|\xi_0|_1 \geq 1$.
 The random walk starts in a state $|u\rangle$ chosen with probability
 $$\frac{\langle u | \xi_0\rangle}{|\xi_0|_1}$$ at time $0$, i.e., the probability of the initial state is proportional to $\langle u | \xi_0\rangle$.
If the random walk is in some state $|u_t \rangle$ at time $t$, then the state of the random walk at time $t+1$ is given by repeating $K$ times the process of picking a random spin and flipping that spin.  Note that we can flip the same spin more than once in a single step of the random walk (indeed, it may be flipped up to $K$ times) although this is unlikely for $K<<\sqrt{N}$.  That is, each step of the random walk we consider here is $K$ steps of a random walk on the Boolean hypercube.
Let $\expec$ denote an expectation value of this random walk.
Then, we have
\begin{eqnarray}
\label{walk1}
\langle \pplus | \phi_{0,1} \rangle &=& 2^{-N/2} \cdot |\xi_0|_1 \cdot \sum_{t=0}^{\infty} B^t \expec\Bigl[\prod_{m=1}^t \frac{1}{E'_{u_{m}}-E_{0,1}}\Bigr] \\ \nonumber
& \geq &
2^{-N/2}  \sum_{t=0}^{\infty} B^t \expec\Bigl[\prod_{m=1}^t \frac{1}{E'_{u_{m}}-E_{0,1}}\Bigr]
\end{eqnarray}
where the random walk has a sequence of states $u_0,u_1,\ldots,u_t$ and
where the inequality uses the fact that $|\xi_0|_1 \geq 1$.

There is one key difference to Ref.~\onlinecite{spa}: we no longer have to condition on the random walk not ``returning".  That is, we now include {\it all} possible sequences of states, rather than just including sequences of states for which the walk does not return to a ground state.
This simplification is why we modified the Brillouin-Wigner perturbation theory.

We have
\be 
\expec\Bigl[\prod_{m=1}^t \frac{1}{E'_{u_{m}}-E_{0,1}}\Bigr]\geq 
\prod_{m=1}^t \frac{1}{\expec\Bigl[E'_{u_{m}}-E_{0,1}\Bigr]},
\ee
by log-convexity of the inverse (see lemma 3 of Ref.~\onlinecite{spa}).
So,
\begin{eqnarray}
\langle \pplus | \phi_{0,1} \rangle &\geq& 2^{-N/2} \sum_{t=0}^{\infty} B^t  \prod_{m=1}^t   \frac{1}{\expec\Bigl[E'_{u_{m}}-E_{0,1}\Bigr]}.
\end{eqnarray}

We have $E'_u\leq E_u+1/2$.
As in Ref.~\onlinecite{spa},
we have
$$\expec[E_{u_m}]\leq (1-\frac{2DmK}{N})E_0.$$  Here we use the fact that for any initial state $u$ with $\langle u | \xi_0\rangle>0$
we have $E_u=E_0$.
Hence,
\begin{eqnarray}
\langle \pplus | \phi_{0,1} \rangle &\geq & 2^{-N/2} \sum_{t=0}^{\infty}B^t  \prod_{m=1}^t   \frac{1}{
\frac{1}{2}+E_0-E_{0,1}+\frac{2DmK}{N} |E_0|}.
\end{eqnarray}

Suppose that $E_{0,1}\geq E_0-1/2$.
Then, for $t\geq 1$,
\begin{eqnarray}
\label{suppose}
&&
\prod_{m=1}^t   \frac{1}{
\frac{1}{2}+E_0-E_{0,1}+\frac{2DmK}{N} |E_0|}
\\ \nonumber
&\geq &
\prod_{m=1}^t   \frac{1}{
1+\frac{2DmK}{N} |E_0|}
\\ \nonumber
& \geq &
\Bigl(\frac{1}{\frac{2DmK}{N} |E_0|}\Bigr)^t \cdot \frac{1}{t!} \cdot
\frac{1}{1+\frac{1}{\frac{2DK}{N} |E_0|}} \cdot
\frac{1}{1+\frac{1}{2}\cdot \frac{1}{\frac{2DK}{N} |E_0|}} \cdot \ldots
\frac{1}{1+\frac{1}{t}\cdot \frac{1}{\frac{2DK}{N} |E_0|}} 
\\ \nonumber
&\geq &
\Bigl(\frac{1}{\frac{2DmK}{N} |E_0|}\Bigr)^t \cdot \frac{1}{t!} \cdot
\exp\Bigl(-\frac{1}{\frac{2DK}{N} |E_0|}\Bigr) \cdot
\exp\Bigl(-\frac{1}{2}\cdot \frac{1}{\frac{2DK}{N} |E_0|}\Bigr) \cdot \ldots
\exp\Bigl(-\frac{1}{t}\cdot \frac{1}{\frac{2DK}{N} |E_0|}\Bigr)
\\ \nonumber
&\geq &
\Bigl(\frac{1}{\frac{2DmK}{N} |E_0|}\Bigr)^t \cdot \frac{1}{t!} \cdot
\exp\Bigl(-\frac{\log(t)+1}{\frac{2DK}{N} |E_0|}\Bigr).
\end{eqnarray}

We have
$\sum_{t=0}^\infty B^t
\Bigl(\frac{1}{\frac{2DmK}{N} |E_0|}\Bigr)^t \cdot \frac{1}{t!}=
\exp(\frac{BN}{2DK|E_0|})$.
Further,
 for $\frac{BN}{2DK|E_0|}$ at most polynomially large, the only non-negligible terms in this series expansion
have $t$ at most polynomially large.
For these terms, we have 
$$\exp\Bigl(-\frac{\log(t)+1}{\frac{2DK}{N} |E_0|}\Bigr)=\exp\Bigl(-\frac{\mO(\log(N))N}{2DK|E_0|}\Bigr).$$
Hence,
\begin{eqnarray}
\langle \pplus | \phi_{0,1} \rangle 
&\geq &
2^{-N/2} \exp\Bigl(\frac{BN}{2DK|E_0|}\Bigr)\exp\Bigl(-\frac{\mO(\log(N))N}{2DK|E_0|}\Bigr),
\end{eqnarray}
and so for
$B=\omega(\log(N))$ we have
\begin{eqnarray}
\label{overlap}
\langle \pplus | \phi_{0,1} \rangle 
&\geq &
2^{-N/2} \exp\Bigl[\frac{BN}{2DK|E_0|}\cdot (1-o(1)) \Bigr].
\end{eqnarray}

This completes the proof of lemma \ref{ovlemma}.

\subsection{Proof of Theorem \ref{Qgood}}
\label{pt}
We now prove theorem \ref{Qgood}.

We first need the following general result, lemma \ref{gen} below.
Consider a Hamiltonian $H$ with a block matrix structure defined by
\be
H=\begin{pmatrix}
\mA & \mB \\
\mB^\dagger & \mC
\end{pmatrix}.
\ee
Assume that $\mA$ has all its eigenvalues upper bounded by some $E_A^{max}$ and $\mC$ has all its eigenvalues lower bounded
by some $E_C^{min}$ with
$E_C^{min}>E_A^{max}$.
Let $\mA$ be an $n_0$-by-$n_0$ matrix.
For application to our problem, we will take $\mA$ to be the Hamiltonian $P H_1 P$ restricted to the range of $P$ and take $\mC$ to be the Hamiltonian $Q H_1 Q$ restricted to the range of $Q$ and take $\mB$ to be the
matrix $P H_1 Q$ restricted to mapping from the range of $Q$ to the range of $P$.
However, for now we will work in generality and derive some results about eigenvalues and overlaps of an arbitrary such $H$.
We show that
\begin{lemma}
\label{gen}
\begin{itemize}
\item[1.] The $n_0$ lowest eigenvalues of $H$ are less than or equal to $E_A^{max}$.  The remaining eigenvalues are all greater than or equal to $E_C^{min}$.

\item[2.] Let the smallest eigenvalue of $\mA$ equal $E_A^{min}$.
Then, the smallest eigenvalue of $H$ is greater than or
equal to the smallest eigenvalue of the $2$-by-$2$ matrix
$$\begin{pmatrix}
E_A^{min} & \Vert \mB \Vert \\
\Vert \mB \Vert & E_C^{min}\end{pmatrix},$$
which is
greater than or equal to
$$E_A^{min}-\frac{\Vert \mB \Vert^2}{E_C^{min}-E_A^{min}}.$$

\item[3.] Let $P$ be the block matrix
\be
P=\begin{pmatrix} I & 0 \\ 0 & 0 \end{pmatrix}.
\ee
Then, for any $\psi$ in the eigenspace of $H$ with eigenvalue less than or equal to $E_A^{max}$ with $|\psi|=1$, i.e., for any $\psi$ in the eigenspace of the $n_0$ smallest eigenvalues of $H$, we have that
\be
\langle \psi | P | \psi \rangle \geq \sqrt{1-\frac{\Vert \mB \Vert^2}{(E_C^{min}-E_A^{max})^2}}.
\ee
\end{itemize}
\begin{proof}
Define the Green's function $G(\omega)=(\omega-H)^{-1}$.
Write $G$ as a block matrix
$$G(\omega)=\begin{pmatrix} G_{00}(\omega) & G_{01}(\omega) \\ G_{10}(\omega) & G_{11}(\omega) \end{pmatrix}.$$
We have
\begin{eqnarray}
G_{00}(\omega)&=& \Bigl(\omega-\mA -\Sigma(\omega)\Bigr)^{-1},
\end{eqnarray}
where
\be
\Sigma(\omega)=\mB (\omega-\mC)^{-1} \mB^\dagger.
\ee

For $\omega<E_C^{min}$, the matrix $\Sigma(\omega)$ is negative semi-definite.
Hence, $G_{00}(\omega)$ does not have any poles in the interval $E_A^{max}<\omega<E_C^{min}$.
Hence, if $H$ has an eigenvalue in this interval, then the corresponding eigenvector has vanishing amplitude on the first block; however, any such eigenvector has eigenvalue equal to an eigenvalue of $\mC$, so no such eigenvector exists.

Thus, all eigenvalues of $H$, for any $v$ are contained in $(-\infty,E_A^{max}] \cup [E_C^{min},\infty)$.
To prove item 1, we use continuity in $\mB$: if $\mB=0$, there are exactly $n_0$ eigenvalues of $H$ in the interval $(-\infty,E_A^{max}]$ and this number does not change as $\mB$ changes, so for any choice of $\mB$ there are exactly $n_0$ eigenvalues of $H$ in the interval $(-\infty,E_A^{max}]$.

To prove item 2, let $\psi$ be an eigenvector with smallest eigenvalue for $H$ with $|\psi|=1$.
Let $x=\langle \psi | P | \psi \rangle$.
We have
\be
\langle \psi | H | \psi \rangle \geq E_A^{min} x + E_C^{min} (1-x) - 2\sqrt{x(1-x)} \Vert \mB \Vert.
\ee
Minimizing over $x$ gives the desired result.

To prove item 3, consider a family of Hamiltonians $H_t$ given by
\be
H_t=\begin{pmatrix}
\mA &  t\mB \\
t\mB^\dagger & \mC
\end{pmatrix},
\ee
where $t$ is a real number.

Let $P_t$ project onto the $n_0$ lowest eigenvalues of $H_t$.
Let $\psi_1$ be some eigenvector in the eigenspace of the $n_0$ lowest eigenvalues of $H_1$ with $|\psi_1|=1$.
We construct a family of vectors $\psi_t$ by
\be
\label{crct}
\partial_t \psi_t=
\eta_t=-\Bigl( \int_0^\infty (1-P_t) \exp(-\tau H_t) \mB \exp(\tau H_t)
P_t {\rm d}\tau \Bigr) \psi_t
\ee
We claim that 
 $\psi_t$ is an eigenvector in the eigenspace of the $n_0$ lowest eigenvalues of $H_1$, with $|\psi_t|=1$.
This may be verified by working in an eigenbasis of $H_t$, doing the integral over $\tau$ exactly in that eigenbasis, and comparing to first order perturbation theory.
Let $E_A^{max}(t)$ equal the $n_0$-th smallest eigenvalue of $H_t$ so that $E_A^{max}(0)=E_A^{max}$, and let
$E_C^{min}(t)$ equal the $(n_0+1)$-st eigenvalue of $H_t$.
So, by results above $E_A^{max}(t)\leq E_A^{max}$ and $E_C^{min}(t)\geq E_C^{min}$.
Then,
\be
\Vert (1-P_t) \exp(-\tau H_t) \mB \exp(\tau H_t)
P_t \Vert \leq \Vert \mB \Vert  \exp(-\tau (E_C^{min}-E_A^{max})).
\ee
Hence, by a triangle inequality,
\be
|\partial_t \psi_t| \leq 
\frac{\Vert \mB \Vert}{E_C^{min}-E_A^{max}}.
\ee
So, since $(1-P)\psi_0=0$, we have that $|(1-P)\psi_1| \leq \frac{\Vert \mB \Vert}{E_C^{min}-E_A^{max}}$.
Hence, item 3 follows.
\end{proof}
\end{lemma}

{\it Proof of theorem \ref{Qgood}:}
We now prove theorem \ref{Qgood}, using lemma \ref{gen}, taking
$\mA$ to be the Hamiltonian $P H_1 P$ in the range of $P$ and taking $\mC$ to be the Hamiltonian $Q H_1 Q$ in the range of $Q$ and taking $\mB$ to be the
matrix $P H_1 Q$ from the range of $Q$ to the range of $P$.
We have $E_A^{min}\geq E_0-1/4$ and $E_A^{max}\leq E_0+1/4$ by the assumption that $\Vert P B (X/N)^K \Vert \leq 1/4$.
We have $E_C^{min}\geq E_0+1/2$.

To prove item 1 of theorem \ref{Qgood}, note that a lower bound of $1/2$ on the smallest eigenvalue of $H_1$ outside the range of $P_1$
follows immediately from item 1 of lemma \ref{gen}.
An upper bound of $E_0+1/4$ on the largest eigenvalue of $H_1$ in the range of $P_1$
also follows immediately from item 1 of lemma \ref{gen}.

To prove item 2 of theorem \ref{Qgood}, note that lemma \ref{ovlemma} gives a lower bound on the
inner product
$\langle \pplus | \psi_{0,1} \rangle \geq
2^{-N/2} \exp\Bigl(\frac{BN}{2DK|E_0|} \cdot (1-o(1)) \Bigr)$, under an assumption that
$E_{0,1} \geq E_0-1/2$.
So, to prove item 2 of theorem \ref{Qgood}, it suffices to prove
that $E_{0,1} \geq E_0-1/2$ under the assumptions of theorem \ref{Qgood} and to prove
that $|\psi_{0,1}| = \mO(1)$, i.e., recall that $\psi_{0,1}$ is not normalized, while theorem \ref{Qgood} makes a
claim about the normalzed state $\psi_{0,1}$.

First, we prove $E_{0,1}\geq E_0-1/2$.  From item 2 of lemma \ref{gen} and the assumption that $E_C^{min}=E^Q_{0,1}\geq E_0+1/2$ and that $\Vert P B (X/N)^K \Vert \leq 1/4$, we have
\begin{eqnarray}
E_{0,1}&\geq & E_A^{min}-\frac{(1/4)^2}{1/2} \\ \nonumber
&\geq & E_0-1/4 - 2 \cdot (1/4)^2 \\ \nonumber
&\geq & E_0-1/2.
\end{eqnarray}

To compute $|\phi_{0,1}|$, use
that $|\phi_{0,1}|=|(E_{0,1}-J_0-V)^{-1} \xi_0|$.
Let $\lambda$ be the smallest eigenvalue of $J_0-V$.
Then, $|\phi_{0,1}|\leq |\lambda-E_{0,1}|^{-1}$.
Define matrix $\mAp=\mA+1/2$.
Then,
$\lambda$ is the smallest eigenvalue of the matrix
$$H'=\begin{pmatrix} \mAp & \mB \\ \mB^\dagger & \mC \end{pmatrix}.$$
By item 2 of lemma \ref{gen}, and since $E_C^{min}\geq E_0+1/2 \geq E_A^{min}+1/2$,
$\lambda$ is greater than or equal to the smallest eigenvalue of the matrix
$$\begin{pmatrix}
E_A^{min}+1/2 & \Vert \mB \Vert \\
\Vert \mB \Vert & E_A^{min}+1/2\end{pmatrix},$$
so
$\lambda\geq E_A^{min}+1/2-\Vert \mB \Vert$.
Using the assumption that $\Vert P B (X/N)^K \Vert \leq 1/4$, we have that
$\lambda \geq E_A^{min}+1/4$ so
$|\phi_{0,1}| \leq 4$.

To prove item 3 of theorem \ref{Qgood}, we need to show that
$\langle \psi | P_0 | \psi \rangle = \Omega(1)$ for any $\psi$ in the range of $P_1$.
By item 3 of lemma \ref{gen},
\begin{eqnarray}
\langle \psi | P_0 | \psi \rangle & \geq &\sqrt{1-\frac{\Vert B \Vert^2}{(E_C^{min}-E_A^{max})^2}}
\\ \nonumber
&\geq & \sqrt{1-\frac{(1/4)^2}{1/4}} \\ \nonumber
&\geq & \sqrt{3/4}.
\end{eqnarray}
This completes the proof of theorem \ref{Qgood}.

\subsection{Proof of Theorem \ref{mainconst}}
\label{mainproof}
We now prove theorem \ref{mainconst}.  The proof is the same as the proof of theorem \ref{mainconst} in Ref.~\onlinecite{spa} with the following changes.  First, theorem \ref{Qgood} here replaces theorem 3 in Ref.~\onlinecite{spa}, generalizing that theorem to the case of multiple ground states.

In theorem \ref{Qgood}, we assume that
$\Vert P B(X/N)^K \Vert \leq 1/4$.  Eq.~(\ref{Pbound}) in appendix \ref{glsi} implies that this bound holds if
Eq.~(\ref{Kbound}) holds.

Then, if the assumption $E^{Q}_{0,1}\geq E_0+1/2$ in theorem \ref{Qgood} holds, theorem \ref{Qgood} implies that item 1 of theorem \ref{mainconst} holds.

If the  assumption  $E^{Q}_{0,1}\geq E_0+1/2$ does not hold, then we apply the same techniques of localization in $X$ and tight log-Sobolev inequalities as in Ref.~\onlinecite{spa} to show that item 2 of theorem \ref{mainconst} holds.
There are only three minor modifications needed to the proof here.

First,, lemma 7 of 
Ref.~\onlinecite{spa} assumes a unique ground state.  As sketched in the discussion of Ref.~\onlinecite{spa}, it is possible to remove this assumption at the cost of worst constants.  We review this here in more detail.  We use the following lemma in place of lemma 7 of Ref.~\onlinecite{spa}:
\begin{lemma}
\label{eigenvlemma}
Assume $E^Q_{0,1}<E_0+1/2$.
Then,
there is an eigenvector $\Psi$ of $H_{5/2}=H_Z-(5/2) B (X/N)^K$ with eigenvalue at most $E_0+1/2$ such that $\langle \Psi | B(X/N)^K | \Psi \rangle \geq 1/4$.
\begin{proof}
Note that $Q H_{5/2} Q$ has an eigenvalue smaller than $E_0+1-(5/2)\cdot 1/2\leq E_0-1/4$, with the corresponding eigenvector in the
range of $Q$.
Hence, $H_{5/2}$ has an eigenvalue smaller than $E_0-1/4$, with corresponding eigenvector $\Psi$ such that
$\langle \Psi | B(X/N)^K | \Psi \rangle \geq 1/4$.
\end{proof}
\end{lemma}

Using this eigenvector $\Psi$, one can repeat the localization in $X$ and log-Sobolev inequality calculations of Ref.~\onlinecite{spa}.
The cost is that some factors of $B$ get replaced by $(5/2) B$; indeed, all factors of $B$ in item 2 of theorem \ref{mainconst} in this paper are multiplied by an additional factor of $5/2$ compared to Ref.~\onlinecite{spa}.

The second modification is that since we allow $K$ even, it is possible that we find wavefunctions with a negative expectation value for $X$; more precisely, after localizing in $X$, we find a wavefunction $\Xi$ (see lemma 9 of Ref.~\onlinecite{spa}) supported on an eigenspace of $X$ with eigenvalues in an interval $[X_0-X_{min}/K,X_0+X_{min}/K]$ for $X_0\leq -X_{min}$.  However, in this case, we may multiply the wavefunction by the unitary $Z_1 Z_2 \ldots Z_N$; this unitary commutes with $H_s$ and changes the sign of $X$; then, we may assume that $X_0\geq X_{min}$.

Finally, since the eigenvalues of $H_Z$ need not be integers, we replace the previous definition of $W(E)$ as the number of eigenvalues of $H_Z$ with energy $E$ with a different definition: it is the number of eigenvalues with energy in the interval $[E,E+1)$; this is used in binning when analyzing the entropy.

This completes the proof.

\section{Arbitrary $D=2$ Models}
\label{arbitrary}
In this section, we consider ways in which the assumption $E^Q_{0,1}\geq E_0+1/2$ of theorem \ref{Qgood} can fail.
Of course, we have derived above by minor modifications of the proof of Ref.~\onlinecite{spa} that if $E^Q_{0,1}<E_0+1/2$ then item 2 of theorem \ref{mainconst} holds, i.e., that there are a large number of computational basis states with energy close to the ground state.
However, while item 2 follows from $E^Q_{0,1}<E_0+1/2$, it is weaker than that.  In this section, we more carefully analyze how we can have 
$E^Q_{0,1}<E_0+1/2$.

We begin with a toy model showing one way in which this can occur.  Then, we give some general results showing, for certain cases, that this toy model exemplifies the only way in which this can occur, in a certain sense made precise below.
We show that
\begin{theorem}
\label{speed}
Assume $D=2$ and that all weights in $H_Z$ are chosen from $\{-1,0,+1\}$.
For any constants $c>0$ and $\alpha\in (11/7,2]$, for $|E_0|\geq c N^{\alpha}$,
there are choices of $K,b$ such that at least one of the following holds:
\begin{itemize}
\item[1.] The short path algorithm finds a ground state in expected time
$$\sO\Bigl(2^{N/2-\Omega(\frac{N^{\frac{2}{3}\alpha-\frac{1}{3}}}{\ln(N)})}\Bigr)
.$$

\item[2.] The number of ground states obeys $\log(n_0)=\Omega(N)$ in which case a Grover search finds the ground state
in expected time
$\sO\Bigl(2^{\frac{N}{2} \cdot(1-\Omega(1))}\Bigr).$
\end{itemize}

Also,
for any constants $c>0$ and $\alpha\in (10/7,11/7]$, for $|E_0|\geq c N^{\alpha}$,
there are choices of $K,b$ such that at least one of the following holds:
\begin{itemize}
\item[1.] The short path algorithm finds a ground state in expected time
$$\sO\Bigl(2^{N/2-\Omega(\frac{N^{3 \alpha-4}}{\ln(N)^2})}\Bigr)
.$$

\item[2.] The number of ground states obeys $\log(n_0)=\Omega(N)$ in which case a Grover search finds the ground state
in expected time
$\sO\Bigl(2^{\frac{N}{2} \cdot(1-\Omega(1))}\Bigr).$
\end{itemize}

\end{theorem}
Finally, we return to the toy model and discuss it in more detail and then discuss possible classical algorithms
as well as improvements to the quantum result above.

\subsection{Toy Model}
One simple way to have $E^Q_{0,1}<E_0+1/2$ is exemplified in the toy model of this subsection.  
Before giving the toy model, recall that one possible problem with the adiabatic algorithm  is that one can have a configuration with energy very close to the ground state but with a few ``flippable spins", i.e., spins that can be flipped without changing the energy, so that this configuration reduces its energy when the transverse magnetic field is applied and, for small but nonzero transverse field, the ground state has small overlap with the desired solution of the optimization problem.
The toy model generalizes this to many flippable spins, as we explain.

We have $N$ qubits.  Divide the qubits into two sets, the first with $N_1$ qubits and the second with $N-N_1$ qubits, with $N_1<<N$.
Write $H_Z=H_{11}+H_{12}+H_{22}$ where $H_{11}$ contains interaction terms within the first set, $H_{12}$ contains term coupling the first set to the second set, and $H_{22}$ contains terms within the second set.
For any given configuration of the first set of qubits, we can describe the problem of optimizing over the second set of qubits as optimizing a function that contains quadratic terms from $H_{22}$
as well as linear terms from $H_{12}$.

Suppose that the interactions between the qubits in the second set are negligible, i.e, that there are very few nonzero interaction terms between them so that $\Vert H_{22} \Vert$ is negligible compared to other terms.
Suppose there are two configurations, $C$ and $C'$ of the first set of qubits with the following properties.  First, $H_Z$ has a unique ground state and this ground state has configuration $C$ on the first set of qubits; for this choice $C$, the expectation value of $H_{11}$ is far from $E_0$ but optimizing $H_{12}+H_{22}$ over configurations on the second set gives energy $E_0$.
Second, $C'$ is such that the expectation value of $H_{11}$ is very close to $E_0$ but the linear terms $H_{12}$ that result from that configuration are small enough that the energy is almost independent of the second set of spins (recall that $H_{22}$ is assumed to be negligible), and so that no configuration of the second set of spins gives energy $E_0$.

Then, we can construct a wavefunction with configuration $C'$ on the first set of spins and with the second set of spins polarized in the $+$ direction.  This gives an energy close to $E_0$ and
gives an expectation value of $B (X/N)^K$ equal to $B ((N-N_1)/N)^K$.
For large $K$, we must have $N_1<<N$ in order to have this quantity non-negligible.
That is, in contrast to the case with the adiabatic algorithm where a small number of flippable spins can create problems, here there is only a problem when there are a large number
of flippable spins.

We will return to this toy model after giving general results in the next subsection.  One of the most important terms in the bounds that we give will be contributions from $H_{12}$; we will later give a different toy example where this occurs, giving a large expectation value of $B(X/N)^K$ as well as a large negative expectation value of $H_{12}$.

\subsection{General Results}
Now consider the case $D=2$ and that all weights are chosen from $\{-1,0,+1\}$.
We will show theorem \ref{speed}.

We will give the specific choices of $b,K$ later.  Before giving those choices, we will analyze in general the ways that the assumptions of theorem
\ref{Qgood}
can fail.  The first way is that
$E^Q_{0,1}<E_0+1/2$, and we consider that case now.

We need to recall the following two results.
First, using lemma \ref{eigenvlemma} above to replace lemma 7 of Ref.~\onlinecite{spa} (this leads to extra factors of $5/2$), by lemma 9 of Ref.~\onlinecite{spa}, if $E^Q_{0,1}<E_0+1/2$, then the following holds.
Let $X_{min}=N\cdot (10B)^{-1/K}$.
Then, there is a state $\Xi$ with $|\Xi|=1$ such that
$\Xi$ is supported on an eigenspace of $X$ with
eigenvalues in some interval
$[X_0-X_{min}/K,X_0+X_{min}/K]$
for $X_0\geq X_{min}$ and such that
\be
\label{e2}
\langle \Xi | H_Z -(5/2)B(X/N)^K| \Xi \rangle \leq E_0+1/2+\mO(1) \frac{J_{tot} K^2D^2}{X_{min}^2}.
\ee
Hence, since $\langle \Xi |(5/2)B(X/N)^K| \Xi \rangle\leq (5/2) B$, we have
so that
\begin{eqnarray}
\label{e3}
\langle \Xi | H_Z | \Xi \rangle & \leq &E_0+1/2+\mO(1) \frac{J_{tot} K^2D^2}{X_{min}^2}+(5/2) B.
 \end{eqnarray}

Second, theorem 3 of Ref.~\onlinecite{brandaoarxiv} (see Ref.~\onlinecite{brandao2013product} for published version) shows the following: given any density matrix $\rho$ on $N$-qubits,
there exists a globally separable state $\sigma$ such that
\be
\expec_{i,j} \Vert \rho^{Q_iQ_j} - \sigma^{Q_i Q_j} \Vert_1 \leq 12 \Bigl(\frac{4 \ln(2)}{N}\Bigr)^{1/3}=\mO(N^{-1/3}).
\ee
Here, the expectation value $\expec_{i,j}$ is with respect to a randomly chosen pair of qubits $i,j$ and $\rho^{Q_iQ_j},\sigma^{Q_iQ_j}$ denote reduced density matrices on qubits $i,j$ and $\Vert \ldots \Vert_1$ is the trace norm.
In fact, an extension of this result holds for states $\rho$ with ${\rm tr}(\rho X)$ close to $N$.  The reason for this extension is that the mutual information of a qubit $i$ with the rest of the system is bounded by $-p \ln(p) - (1-p) \ln(1-p)$ with $p={\rm tr}((1-X_i) \rho)/2$, and if ${\rm tr}(\rho X)$ is close to $N$ then for many of the qubits $p$ will be small.  Hence, after some averaging one may show that\cite{aharrow}:
\be
\label{prodapprox}
\expec_{i,j} \Vert \rho^{Q_iQ_j} - \sigma^{Q_i Q_j} \Vert_1 \leq 12 \Bigl(\frac{4 \ln(2)}{N}\Bigr)^{1/3}=\mO(\en^{1/3} N^{-1/3}),
\ee
for
\be
\en \leq -{\rm const.} \times \Bigl(1-{\rm tr}(\rho X/N)\Bigr) \cdot \ln\Bigl(1-{\rm tr}(\rho X/N)\Bigr).
\ee
Without this extension, our proof of theorem \ref{speed} would be restricted to $\alpha$ in the interval $(5/3,2]$.
Remark: we will pick $K=\Omega(\log(N))$ so that $X_{min}$ is $\Omega(N)$.  Indeed, we will pick $K$ sufficiently large that $S=o(1)$.

We let $\rho=|\Xi\rangle\langle \Xi|$ and find a separable state $\sigma$ obeying
Eq.~(\ref{prodapprox}).
Hence, since ${\rm tr}(\rho X)\geq X_{min}-X_{min}/K$, we have
\be
{\rm tr}(\sigma X)\geq X-\Delta_X,
\ee
where
\be
\Delta_X\equiv  (N-X_{min})+X_{min}/K+\mO(\en^{1/3} N^{2/3}).
\ee
The term $\mO(\en^{1/3} N^{2/3})$ arises from the $\mO(\en^{1/3} N^{-1/3})$ error term in Eq.~(\ref{prodapprox}) multiplied by $N$, which is the number of qubits.

Also, since
${\rm tr}(\rho H_Z) \leq 
E_0+1/2+\mO(1) \frac{J_{tot} K^2D^2}{X_{min}^2}+(5/2) B,$
we have
\be
{\rm tr}(\sigma H_Z) \leq E_0 + \Delta_H,
\ee
where
\be
\Delta_H \equiv 1/2+\mO(1) \frac{J_{tot} K^2D^2}{X_{min}^2}+(5/2) B+\mO(\en^{1/3} N^{5/3}).
\ee
The additive factor of $\en^{1/3} N^{5/3}$ arises because for a randomly chosen pair of sites, the average trace norm difference
between $\rho$ and $\sigma$ on that pair of sites is $\mO(\en^{1/3} N^{-1/3})$; there are $\mO(N^2)$ pairs of sites and each
$|J_{ij}|$ is bounded by $1$ in absolute value.
Remark: here we are considering the case $D=2$ so the factor $D^2$ in the definition of $\Delta_H$ is simply a constant (i.e., $4$).

By definition, such a separable state $\sigma$ can be written as 
a convex combination of product states.
Hence, some product state $\tau$ must have
\be
{\rm tr}(\tau X)\geq X-2 \Delta_X
\ee
and
\be
\label{trHzbd}
{\rm tr}(\tau H_Z) \leq E_0 + 2\Delta_H.
\ee

We now consider this particular state $\tau$.  
Define
$$\delta_x=\frac{\Delta_X}{N}.$$
For any $c>1$, the probability that a randomly chosen qubit $i$
has ${\rm tr}(\tau X_i)< 1-2c\delta_X$ is at most $c^{-1}$,
where $X_i$ is the $X$ operator on qubit $i$.  Leaving $c$ unspecified for now (later we will pick $c=(1/8)\delta_x^{-1}$), we divide the qubits into
two sets, labelled $S_1,S_2$.  The second set, $S_2$, will be those qubits $i$ such that ${\rm tr}(\tau X_i)\geq 1-2c \delta_x$
and the first set, $S_1$, will be the remaining qubits.
Let the first set have $N_1$ qubits, so that $N_1\leq c^{-1} N$.

Let
$H_Z=H_{11}+H_{12}+H_{22}$ where $H_{11}$ contains interaction terms within the first set, $H_{12}$ contains terms coupling the first set to the second set, and $H_{22}$ contains terms within the second set.
For any qubit $i$ we have
\be
\label{zbound}
{\rm tr}(\tau Z_i) \leq \sqrt{1-{\rm tr}(\tau X_i)^2}\leq \sqrt{2\Bigl(1- {\rm tr}(\tau X_i)\Bigr)}.
\ee
Hence,
\begin{eqnarray}
\label{tr12bd}
\Bigl| {\rm tr}(\tau H_{12}) \Bigr|&\leq & N_1  \sum_{i\in S_2} |{\rm tr}(\tau Z_i)| \\ \nonumber
& \leq & N_1 \cdot \sum_i  |{\rm tr}(\tau Z_i)|  \\ \nonumber
& \leq & N_1 \cdot N   \sqrt{2\Bigl(1- {\rm tr}(\tau \frac{X}{N})\Bigr)} \\ \nonumber
&\leq & N_1 \cdot N \sqrt{4\delta_x}  \\ \nonumber
&=& 2N_1 \cdot N \sqrt{\delta_x} \\ \nonumber
& \leq & 2 c^{-1} \cdot N^2  \sqrt{\delta_x}.
\end{eqnarray}
where we have use concavity of the square-root.
So,
\be
\label{tr12nabs}
{\rm tr}(\tau H_{12}) \geq -2 c^{-1} \cdot N^2 \sqrt{\delta_x}.
\ee

Next we bound ${\rm tr}(\tau H_{22})$.  Given the state $\tau$, we construct a new product state $\tau'$ as follows.
First, in $\tau'$, all qubits in the first set will be polarized in the $X$ direction so that ${\rm tr}(\tau' H_{11})={\rm tr}(\tau' H_{12})=0$.  Second, the qubits in the second set will have
\be
{\rm tr}(\tau' Z_i)=\frac{{\rm tr}(\tau Z_i)}{\sqrt{4c \delta_x}}.
\ee
By Eq.~(\ref{zbound}), this value of ${\rm tr}(\tau' Z_i)$ is achievable since ${\rm tr}(\tau X_i)\geq 1-2c \delta_x$ for the qubits in this set.
Hence,
\be
{\rm tr}(\tau' H_{22})=\frac{{\rm tr}(\tau H_{22})}{4c\delta_x}.
\ee
Thus, since ${\rm tr}(\tau' H_{22})\geq E_0$,
we have
\be
\label{tr22bd}
{\rm tr}(\tau H_{22}) \geq 4c\delta_x E_0.
\ee

Hence, from Eqs.~(\ref{tr12nabs},\ref{tr22bd}) we have
\begin{eqnarray}
\label{trHZ}
{\rm tr}(\tau H_Z) &=& {\rm tr}(\tau H_{11}) + {\rm tr}(\tau H_{12})+{\rm tr}(\tau H_{22}) \\ \nonumber
& \geq &  {\rm tr}(\tau H_{11}) - 2  c^{-1}  \sqrt{\delta_x}\cdot N^2  +4c\delta_x E_0.
\end{eqnarray}
Let us fix
\be
c=(1/8) \delta_x^{-1}.
\ee 
Then from Eqs.~(\ref{trHzbd},\ref{trHZ}), we have
\be
{\rm tr}(\tau H_{11}) -  16 \delta_x^{3/2} N^2+ \frac{E_0}{2} \leq E_0 + 2 \Delta_H.
\ee
Hence,
\be
{\rm tr}(\tau H_{11}) -  16 \delta_x^{3/2} N^2\leq \frac{E_0}{2} + 2 \Delta_H.
\ee

Clearly
\be
\Bigl| {\rm tr}(\tau H_{11}) \Bigr| \leq N_1^2 \leq 64 \delta_x^2 N^2.
\ee
Thus,
\be
\label{upperbound}
-64 \delta_x^2 N^2 -16 \delta_x^{3/2} N^2- 2\Delta_H \leq \frac{E_0}{2}.
\ee
Hence,
\be
\label{upperboundabs}
64 \delta_x^2 N^2 +16 \delta_x^{3/2} N^2+2\Delta_H \geq \frac{|E_0|}{2}.
\ee

For sufficiently small $\delta_x,\Delta_H$ and sufficiently large $|E_0|$, Eq.~(\ref{upperboundabs}) has no solution.
We choose
$$B=-\frac{E_0}{10}.$$
That is, we fix $b=1/10$.
We have
\begin{eqnarray}
X_{min}&=& N\cdot (10B)^{-1/K} \\ \nonumber
&=& N \cdot E_0^{-1/K} \\ \nonumber
&=& N \exp\Bigl(-(1/K) \ln(|E_0|)\Bigr) \\ \nonumber
&\geq & N -N \frac{\ln(|E_0|)}{K} \\ \nonumber
&\geq & N-N\frac{2\ln(N)}{K}.
\end{eqnarray}
Thus,
\be
\en=\mO(\frac{\ln(N)}{K}\ln\Bigl(\frac{K}{\ln(N)}\Bigr)).
\ee
Hence,
\begin{eqnarray}
\label{dXis}
\delta_X&\leq &\frac{2\ln(N)+1}{K}+\mO(\en^{1/3} N^{-1/3}).
\end{eqnarray}

For sufficiently large $K$ such that $X_{min}\geq N/2$ (thus, $K\geq 4\ln(N)$), we have
\begin{eqnarray}
\label{dHis}
\Delta_H &=& 1/2+\mO(1) \frac{J_{tot} K^2D^2}{X_{min}^2}+(5/2)B+\mO(N^{5/3}) \\ \nonumber
\\ \nonumber
&\leq & 1/2+\mO(1)  K^2D^2+\frac{|E_0|}{4}+\mO(\en^{1/3} N^{5/3}),
\end{eqnarray}
where we use that $J_{tot}\leq N^2$.

Using Eqs.~(\ref{upperboundabs},\ref{dXis},\ref{dHis}) we find that
\be
\mO(\delta_X^2 N^2)+\mO(\delta_X^{3/2} N^2) +\mO(1)+\mO(K^2) +\mO(\en^{1/3} N^{5/3}) \geq \frac{|E_0|}{4},
\ee
where the big-O notation considers asymptotic dependence on $N,K$.  Multiplying the equation through by $4$ and dropping the term
$\mO(\delta_X^2 N^2)$ which is asymptotically negligible compared to $\delta_X^{3/2} N^2$ we find
\be
\label{hassoln}
\mO(\frac{\ln(N)^{3/2} N^2}{K^{3/2}})+\mO(\frac{\ln(N)^{1/2} N^{3/2}}{K^{1/2}}\ln\Bigl(\frac{K}{\ln(N)}\Bigr)^{1/2})
+\mO(K^2) +\mO(\frac{\ln(N)^{1/3} N^{5/3}}{K^{1/3}}\ln\Bigl(\frac{K}{\ln(N)}\Bigr)^{1/3}) \geq |E_0|,
\ee
where the sum of the first two terms on the left-hand side is a bound on $\mO(\delta_X^{3/2} N^2)$.
In fact the second term on the left-hand side is asymptotically negligible compared to the last term and may also be dropped.
For $K=\Omega(\ln(N) N^{\bex})$ for
\be
\bex=\frac{4}{3}-\frac{2}{3}\alpha,
\ee
the first term is $\mO(N^{\alpha})$, i.e., it is of the same magnitude as $|E_0|$.
For $K=\Omega(\ln(N)^2 N^\nu)$ for
\be
\nu=5-3\alpha,
\ee
the last term is $\mO(N^{\alpha})$.
Finally, for $K=\mO(N^{\alpha/2})$, the term $\mO(K^2)$ is $\mO(N^\alpha)$.

Considering the range of $\alpha$ for which these different constraints on $K$ become asymptotically most important, we find that
for any constants $c>0$ and $\alpha\in (11/7,2]$, for $|E_0|\geq c N^{\alpha}$,
there is some constant $C>0$ such that for
$K\in [C \ln(N) N^{\bex},2C\ln(N) N^{\bex}]$,
Eq.~(\ref{hassoln}) has no solution
for all sufficiently large $N$,
as can be seen by considering the leading order asymptotic behavior of Eq.~(\ref{hassoln}) at large $N$.
In this range, the last term $\mO(\frac{\ln(N)^{1/3} N^{5/3}}{K^{1/3}}\ln\Bigl(\frac{K}{\ln(N)}\Bigr)^{1/3})$ is negligible.
Remark: we have given an upper bound $2C\ln(N) N^{\bex}$ on $K$ because for sufficiently large $K$ the term $\mO(K^2)$ may become the dominant term on the left-hand side
of Eq.~(\ref{hassoln}); however, any upper bound by a constant multiple of $\ln(N) N^{\bex})$ would suffice.
Also, for any
for any constants $c>0$ and $\alpha\in (10/7,11/7]$, for $|E_0|\geq c N^{\alpha}$,
there is some constant $C>0$ such that for
$K\in [C \ln(N)^2 N^{\nu},2C\ln(N) N^{\bex}]$,
Eq.~(\ref{hassoln}) has no solution
for all sufficiently large $N$.

Thus, for any $K$ in this range and for this choice of $b$, we cannot have $E^Q_{0,1}<E_0+1/2$.
The other assumption of theorem \ref{Qgood} is that $\Vert P B(X/N)^K \Vert \leq 1/4$.
By Eq.~(\ref{XNKbnd}), this holds if
\be
B \cdot \cfn\Bigl(\frac{\log(n_0)}{N}+\frac{(K+1/2)\log(N)+1}{N}\Bigr)^{K} \leq 1/4.
\ee
For $\alpha\in (11/7,2]$, we have $\bex\in[0,2/7)$.
For $\alpha\in(10/7,11/7]$, e have
$\nu \in [2/7,5/7)$.
Hence,
$\frac{(K+1/2)\log(N)+1}{N}=o(1)$.
Thus, $\Vert P B(X/N)^K \Vert \leq 1/4$ for all sufficiently large $N$ unless
$\log(n_0)$ is $\Omega(N)$.

For any $K$ in this range and for this choice of $b$,
the exponential
of $\frac{bN}{2DK}$ gives the speedup show in
theorem \ref{speed}.  This completes the proof.

\subsection{Return to Toy Model}
It is interesting that in Eq.~(\ref{upperbound}), the most significant dependence on $\delta_x$ for small $\delta_x$ is the term $ -8 \delta_x^{3/2} N^2$.  This term arises from $H_{12}$.
Consider now the following toy model: there is a ferromagnetic interaction between all spins in $S_1$, i.e.,
$$H_{11}=-\sum_{i,j\in S_1} Z_i Z_j.$$
We consider a product state $\tau$ in which all the spins in $S_1$ have $Z_i=+1$ (or, equivalently, all have $Z_i=-1$).  This gives indeed the most negative possible value of ${\rm tr}(H_{11} \tau)$.  We let
$$H_{12}=-\sum_{i\in S_1,j\in S_2} Z_i Z_j.$$

We choose $H_{22}$ to be a weak anti-ferromagnetic interaction.  Let us explain what we mean by ``weak".  If we had allowed the $J_{ij}$ to be arbitrary real numbers, then we could choose $H_{22}=j \sum_{i,j\in S_2} Z_i Z_j$ for some small $j>0$.  However, since we restrict the $J_{ij}$ to be chosen from $\{-1,0,+1\}$, we can choose $H_{22}$ to equal $\sum_{i,j \in S_2} J_{ij} Z_i Z_j$ where for $i,j\in S_2$ we have $J_{ij}$ either equal to $0$ or $+1$ and only some small fraction of pairs $i,j$ (chosen randomly) have $J_{ij}=+1$.  We will explain below why we choose such an anti-ferromagnetic interaction.

Then, we choose $\tau$ so that any spin $i\in S_2$ has expectation ${\rm tr}(\tau X_i)=1-\delta_x$ so that ${\rm tr}(\tau Z_i)$ is proportional to $\delta_x^{1/2}$.  We choose ${\rm tr}(\tau Z_i)$ positive for all $i\in S_2$.
This means that the antiferromagnetic interaction term $H_{22}$ has positive expectation value, but the interaction $H_{12}$ has negative expectation value proportional to $\delta_x^{3/2}$.
The interaction term $H_{12}$ gives an effective magnetic field on each spin in $S_2$ since the spins in $S_1$ are polarized in the $Z$ direction, but this field is enough only to slightly polarize the spins in $S_2$.

Now we can explain why we choose an anti-ferromagnetic $H_{22}$.  Consider the state $\tau$.  Define a state $\tau''$
as follows.  The spins in $S_1$ will still have $Z_i=+1$ in $\tau''$.  The spins in $S_2$ will have
\be
{\rm tr}(\tau'' Z_i)=C {{\rm tr}(\tau Z_i)},
\ee
for some constant $C$.  From the assumption that the spins in $S_2$ have ${\rm tr}(\tau Z_i)$ proportional to $\delta_x^{1/2}$, for small $\delta_x$ we can choose $C$ large (indeed, we can choose $C$ proportional to $\delta_x^{-1/2}$). We have ${\rm tr}(\tau'' H_{12})=C {\rm tr}(\tau H_{11})$ so increasing $C$ makes the energy of $\tau''$ more negative.
For consistency however, there cannot be a choice of $C$ that makes the energy smaller than $E_0$.
This, then is why we added an anti-ferromagnetic $H_{22}$: adding this term gives a positive contribution to
the energy proportional to $C^2$ and so we can balance these terms so that the optimal choice of $C$ is at $C=1$, i.e., so that the energy of $\tau''$ cannot be reduced below the energy of $\tau$.

This toy model has a state with a large expectation value of $(X/N)^K$ and a large negative expectation value of $H_{12}$.

\subsection{Classical Algorithms}
Certainly, since we have given only a very slight speedup in this case, it is possible that applying a Grover speedup to some classical algorithm may achieve similar performance.  Let us give one simple idea, which, however, does not seem to quite work (some refinement of this idea may work).
Define
\be
F_i=\sum_{j\neq i} J_{ij} Z_j.
\ee
We have
\be
H_Z=\frac{1}{2} \sum_i Z_i F_i.
\ee
Hence, some $i$ must have $|F_i| \geq 2|E_0|/N$.

If, for that given $i$, there are $M$ choices of $j\neq i$ such that $J_{ij}\neq 0$, then the number of computational basis states which give $|F_i| \geq 2E_0/N$ is equal to
$$n_{choice} \equiv \sum_{F \geq 2 |E_0|/N} 2^{N-1-M} {M \choose \frac{M+F}{2}}.$$
Note that $M\leq N-1$.

Thus, for any $c>0$ and any $\alpha$, if $|E_0| \geq cN^{\alpha}$, then for some $i$ we have $F_i=\Omega(N^{\alpha-1})$, so that
\be
n_{choice} \leq 2^{N-\Omega(N^{2\alpha-3})}.
\ee
Consider then the classical algorithm which considers each of the $N$ possible choices of $i$, and then
does a brute force search over all computational basis states with $|F_i| \geq 2E_0/N$.
This can be done in time
$$\sO\Bigl(2^{N-\Omega(N^{2\alpha-3})}\Bigr).$$
However, even applying a Grover speedup to this algorithm is still slower for any $\alpha<2$ than the short-path algorithm.
Indeed, it does not give rise to any nontrivial speedup for $\alpha\leq 3/2$.
It may be possible though to further speedup this algorithm by exploiting the constraint that many choices of $i$ must have large $|F_i|$.

\section{Random Models}
\label{random}
We now briefly discuss the case of random models.  
We emphasize that the remarks in this section are all heuristic; they are mostly intended to give some references to relevant literature and to explain some physics intuition.

For random choices of the $J_{ij}$, we would like to argue that with high probability 
item 2 of theorem \ref{mainconst} wll not hold and so item 1 holds, allowing polynomial speedup compared to Grover search.
Roughly, if $\log(W(E))$ is bounded by a polynomial in $E-E_0$ (in particular for small $E-E_0$) and $\log(n_0)=\mO(N^\gamma)$ for some $\gamma<1$, then item 2 will not hold for some choice of $K,b$ which depends only on these polynomials and not on $N$.

More precisely: note that item 2 has the condition that $\log(W(E))\geq F^{-1}(E)-\mO(\log(N))$.  If we ignore the $\mO(\log(N))$ term, and instead just assume that $\log(W(E)) \geq F^{-1}(E)$, and ignore the term
$\mO(1)\frac{J_{tot}K^2D^2}{X_{min}^2}$ in Eq.~(\ref{FSdef}), then we obtain a condition that $B(\tau(\log(W(E))/N))^K \geq E-E_0$.
We will justify ignoring this term in more detail below.
From the behavior of $\tau(\sigma)$ at small $\sigma$, which is polynomial in $\sigma$ up to logarithmic factor, this gives a polynomial inequality relating between $\log(W(E))$ and $E-E_0$ (up to logarithmic factors) with the power
depending on $K$.  Thus, by adjusting $B,K$, if $\log(W(E))$ has a polynomial dependence on $E-E_0$, we can ensure that this inequality does not hold.

We briefly review some results along this line.

If we have $D=2$ and the quantities $J_{ij}$ are chosen randomly from a Gaussian distribution with unit variance, then there is a very detailed understanding
of the properties of the model.  This model is called the Sherrington-Kirkpatrick model\cite{kirkpatrick1975solvable}.  The Parisi solution\cite{parisi1979infinite,parisi1983order}  gives a detailed understanding of the properties of the model with many results rigorously proven\cite{guerra2003broken,talagrand2006parisi}.  In particular, the entropy $\log(W(E))$ has the desired power-law dependence\cite{crisanti2002analysis,pankov2006low}.
One has $\log(W(E))\sim (E-E_0)^{2/3}$.  This implies $\log(W(E)) \sim N \cdot ((E-E_0)/|E_0|)^{2/3}$ since $|E_0|\sim N^{3/2}$ with high probability; the references given use a different normalization of the $J_{ij}$ so that $|E_0| \sim N$ in those references, but using the normalization here, we find the given power-law for $\log(W(E))$.  This dependence of $\log(W(E))$ strongly suggests that we have a scaling function that $\log(W(E))/N$ is proportional to some function of $(E-E_0)/|E_0|$ over the full range of energies, with the given power-law dependence at small $E-E_0$.
However, we remark
that it is not clear if this behavior holds only for $E-E_0=\Theta(E_0)$ or for all choice of $E$, i.e., whether it includes $E-E_0<<E_0$.  

For the case of Gaussian $J_{ij}$, let $\Delta$ be the energy gap.  If $\Delta$ is only polynomially small (this does not seem to be known in the literature) then we can scale the $J_{ij}$
so that $\Delta=1$.  Then, $J_{tot}$ will be only polynomially large.
The magnitude of the term
$\mO(1)\frac{J_{tot}K^2D^2}{X_{min}^2}$ will depend on $K$ and on how large $J_{tot}$ is; we do not have an estimate for this.

Now consider the case where the $J_{ij}$ to be chosen from a discrete distribution with integer values.  For example, one can use the so-called ``$\pm J$" model where we choose each one independently to be $+1$ with probability $1/2$ and $-1$ with probability $1/2$.
In this case, it seems likely that the same behavior of $\log(W(E))$ would hold, but we do not know a reference.
Another concern is that the power-law dependence of $\log(W(E))$ is only shown for $E-E_0=\Theta(E_0)=\Theta(N^{3/2})$, while we need these bounds also for $E-E_0=1,2,\ldots$.
However, it seems likely that the bounds would hold there too.

On the other hand, one advantage of a discrete distribution of the $J_{ij}$ is the automatically we have a gap $1$ between ground and first excited state without rescaling the couplings.
For the $\pm 1$ model, we have $J_{tot}=\Theta(N^2)$, and so
the term
$\mO(1)\frac{J_{tot}K^2D^2}{X_{min}^2}$ will be $\Theta(N^{2/K})$.
Unfortunately, for any fixed $K$ this is much larger than $1$.  However, the term $J_{tot} K^2 D^2/X_{min}^2$ arose from a worst case treatment of certain error terms in Ref.~(\onlinecite{spa}).  For a random model, although $J_{tot}=\Theta(N^2)$, with high probability $E_0=\Theta(N^{3/2})$; this behavior of $E_0$ is known for the Gaussian model and likely true for the $\pm 1$ model.
We expect that a more accurate treatment will make the error terms of order
$\mO(1)\frac{E_0 K^2D^2}{X_{min}^2}$ which is $o(1)$ for sufficiently large $K$.

One might also consider the case $D>2$.  This is also studied in the physics literature, where it is called the ``$p$-spin model"\cite{derrida1981random}.  The letter $p$ rather than $D$ is used in the physics literature for this case.
The solution of the model for $D>2$ is simpler than that for $D=2$.  For $D>2$, it displays what is called ``one-step replica symmetry breaking"\cite{gardner1985spin}  Roughly, this means
that at low temperatures, the dominant contributions to the free energy come from several distinct basins; each basin corresponds to some region in the Boolean hypercube, but the basins are separated from each other by a large Hamming distance.  In this case, we can separate the entropy at low temperature into two distinct contributions: one coming from the probability distribution over basins and one combing from the entropy within a given basin.  However, since the basins are separated from each other by such a large distance, the term $X^K$ for $K=\mO(1)$ may not able to move from one basin to another.
This may allow one to use even a smaller value of $K$ and a large value of $B$; even if for this choice of $K,B$ it is possible to satisfy item 2. of theorem \ref{mainconst}, it may still satisfy the
assumptions of theorem \ref{Qgood}.  That is, the density of states condition is sufficient but not necessary to achieve the needed bound on $E^Q_{0,1}$ and it may be sufficient just to obey the density of states conditions considering the number of states in a {\it single basin}.

\section{Discussion}
We have generalized some of the result on the short-path algorithm, in particular to degenerate ground states.
We have more carefully analyzed ways in which this algorithm can fail, showing that (for large enough $E_0$) the only possibility involves
having a large number of flippable spins in a sense that we have made precise.  From this, some very weak speedup can be derived for arbitrary models; we expect that larger speedups occur for random models.

\appendix
\section{Generalized Log-Sobolev Inequality and bounds on $\Vert P (X/N)^K \Vert$}
\label{glsi}
In this section, we prove bounds on the expectation value $\langle \psi | (X/N)^{2K} | \psi \rangle$ for an arbitrary
state $\psi$, and an arbitrary positive integer $K$.  
The bounds are proven in terms of various entropies.  If we restrict to $\psi$ supported on a set of computational basis states with cardinality $n_0$, this will give a bound on $\langle \psi | (X/N)^{2K} | \psi \rangle$ in terms of $n_0$.

The main result expressed in terms of entropies defined below is Eq.~(\ref{genineq}) below.
This equation depends upon a sequence of entropies $S_i$ defined in Eq.~(\ref{Sidef}).
To define these entropies, for any quantum state $\phi$, define $\sz(\phi)$ to be the entropy of the probability distribution of measurement outcomes when measuring the state in the computational basis; for example, $\sz(\pplus)=N$.

If $\psi$ is supported on a set of cardinality $n_0$, then the main result is Eq.~(\ref{genineqbasis}) below, which in turn implies the following slightly looser (but more easily expressed) bound:
\begin{eqnarray}
\label{loose}
\langle \psi | (X/N)^{2K} | \psi \rangle
&\leq &
\cfn\Bigl(\frac{\log(n_0)+(K+1/2)\log(N)+1}{N}\Bigr)^{2K} \\ \nonumber
&=&
\cfn\Bigl(\frac{\log(n_0)}{N}+\mO(\frac{\log(N)}{N})\Bigr)^{2K}.
\end{eqnarray}
This implies the bound (\ref{XNKbnd}) claimed earlier which we repeat here:
\be
\label{Pbound}
\Bigl| (X/N)^{K} | \psi \rangle \Bigr|
\leq
\cfn\Bigl(\frac{\log(n_0)}{N}+\frac{(K+1/2)\log(N)+1}{N}\Bigr)^{K}.
\ee

First, we will recall the bound on $\langle \psi | (X/N) | \psi \rangle$ in terms of $\sz(\psi)$, and then we will show how this bound implies
bounds on  $\langle \psi | (X/N)^{2K} | \psi \rangle$.  We remark that similar bounds can be proven on
$\langle \psi | (X/N)^{L} | \psi \rangle$ for odd $L$ but we will not need those bounds.

The bound on $\langle \psi | (X/N) | \psi \rangle$  follows from the tight log-Sobolev inequality of
Ref.~\onlinecite{samorodnitsky2008modified}.
As shown in Ref.~\onlinecite{spa}, this inequality implies the following lemma:
\begin{lemma}
\label{lsi2}
Let $S(x)=-x \log(x)-(1-x) \log(1-x)$ be the binary entropy function (we use $S$ rather than $H$ to avoid confusion with the Hamiltonian $H$).
Let
\be
\cfn(\sigma)=2 \sqrt{S^{-1}(\sigma) \Bigl(1-S^{-1}(\sigma)\Bigr)}\Bigr).
\ee
(The inverse of $S$ may be chosen arbitrarily so long as the same inverse is chosen in both locations.)
Then,
\be
\label{sx}
\cfn\Bigl(\frac{\sz(\psi)}{N}\Bigr)\geq  \frac{ \langle \psi | X | \psi \rangle}{N}.
\ee
\end{lemma}

Let us call $\sz(\psi)/N$ the {\it entropy density}.
The inequality has a simple interpretation: for given $\sz(\psi)$, the expectation value $\langle \psi | X | \psi \rangle$
is maximized by taking
a product state, with all qubits in the same state $a |0\rangle + b |1\rangle$, with non-negative $a,b$ chosen so that the qubit, after measurement in the computational basis, has an entropy equal to the entropy density $\sz(\psi)/N$.
Thus, the inequality is tight since the upper bound is achieved by this product state.

Unfortunately, no good bound can be proven for $\langle \psi | (X/N)^{2K} | \psi \rangle$ just in terms of $\sz(\psi)$.
As an example, consider a state $|u\rangle + N^{-1/2} \pplus$ for any computational basis state $|u\rangle$.
The entropy (after appropriately normalizing the state) is $\mO(1)$, but 
$\langle \psi | (X/N)^{2K} | \psi \rangle$ is roughly $1/N$ for all $K>0$.

We now give an upper bound on $\langle \psi | (X/N)^{2K} | \psi \rangle$ in terms of several different entropies.
Define a sequence of states $\psi_i$ for $i=0,\ldots,K$ by
\begin{eqnarray}
\psi_0=\psi, \\ \nonumber
i>0 \; \rightarrow \;
\psi_i = \frac{X \psi_{i-1}}{|X\psi_{i-1}|}.
\end{eqnarray}
Then,
\begin{eqnarray}
\langle \psi | (X/N)^{2K} | \psi \rangle &=&
\prod_{i=0}^{K-1} \frac{\langle \psi | (X/N)^{2i+2} | \psi \rangle}{\langle \psi | (X/N)^{2i} | \psi \rangle}
\\ \nonumber
&=&
\prod_{i=0}^{K-1} |\langle \psi_{i+1} | X/N | \psi_i \rangle|^2.
\end{eqnarray}

Let
\be
\label{Sidef}
S_i=\sz(\psi_i).
\ee
We maximize $\langle \psi | (X/N)^{2K} | \psi \rangle$ for a given sequence of $S_i$.
Let the ``even" computational basis states be those with even Hamming weight, and let the ``odd" computational basis states be those with odd Hamming weight.
We may assume that $\psi$ is supported on states with even Hamming weight.
Hence, we may assume that $\psi_i$ is supported on even states for even $i$ and odd states for odd $i$.
Thus, the expectation value 
$\langle \psi_{i+1} | X/N | \psi_i \rangle|$ is equal to $\langle (1/\sqrt{2}) (\psi_{i+1} + \psi_i) | X/N| (1/\sqrt{2})(\psi_{i+1} + \psi_i)  \rangle$.
By lemma \ref{lsi2},
we have
\begin{eqnarray}
\langle \psi_{i+1} | X/N | \psi_i \rangle| &\leq &
\cfn\Bigl(\frac{\sz((1/\sqrt{2}) (\psi_{i+1} + \psi_i))}{N}\Bigr)
\\ \nonumber
& \leq &
\cfn\Bigl(\frac{\frac{S_i+S_i+1}{2}+1}{N}\Bigr).
\end{eqnarray}

Hence, we have
\be
\label{genineq}
\langle \psi | (X/N)^{2K} | \psi \rangle
\leq
\prod_{i=0}^{K-1} \cfn\Bigl(\frac{\frac{S_i+S_i+1}{2}+1}{N}\Bigr)^2.
\ee

If the state $\psi$ is supported on a set of computational basis states with cardinality $n_0$, then
$\psi_i$ is supported on a set with cardinality at most $n_0\cdot N^i$.
Hence,
\be
S_i\leq \log(n_0)+i\cdot \log(N).
\ee
Hence, in this case,
\be
\label{genineqbasis}
\langle \psi | (X/N)^{2K} | \psi \rangle
\leq
\prod_{i=0}^{K-1} \cfn\Bigl(\frac{\log(n_0)+(i+1/2)\log(N)+1}{N}\Bigr)^2.
\ee
From this, Eq.~(\ref{loose}) follows easily.

While the bounds \ref{genineqbasis},\ref{loose} are probably not tight, we can give one simple example to illustrate that they are not far from optimal.  Consider $n_0=1$.  Consider the asymptotic behavior for fixed $K$ at large $N$.
From the asymptotic behavior of $\cfn(\sigma)$ for small $\sigma$, for $\sigma=\Theta(\log(N)/N)$, we have $\cfn(\sigma)=\Theta(1/\sqrt{N})$ and hence
the bound gives $\langle \psi | (X/N)^{2K} | \psi \rangle \leq \mO(1/N^{K})$ which agrees with the exact result for $n_0=1$ in Ref.~\onlinecite{spa}.

 \bibliography{sg-ref}
\end{document}